\documentclass[10pt,aps,prd,twocolumn,superscriptaddress,floatfix,amsmath,amssymb,amsfonts,longbibliography,nofootinbib]{revtex4-2}

\usepackage[english]{babel}
\usepackage{graphicx}
\usepackage{dcolumn}
\usepackage{bm}

\usepackage{physics}
\usepackage{amsmath,amssymb,mathtools}
\usepackage{booktabs}
\usepackage{tikzsymbols}
\usepackage[utf8]{inputenc}
\usepackage[T1]{fontenc}

\usepackage{tikz}
\usetikzlibrary{arrows.meta, positioning}
\usepackage{lipsum}
\usepackage{hyperref}

\hypersetup{
    colorlinks=true,
   linkcolor=blue,
    filecolor=red,      
    urlcolor=cyan,
}

\begin{document}
\title{Quantum Geometric Phases as a New Window on Gravitational Waves}

\author{Partha Nandi}
\email{pnandi@sun.ac.za}
\affiliation{Department of Physics, University of Stellenbosch, Stellenbosch-7600, South Africa}
\affiliation{National Institute for Theoretical and Computational Sciences (NITheCS), South Africa}

\author{Frederik G. Scholtz}
\email{fgs@sun.ac.za}
\affiliation{Department of Physics, University of Stellenbosch, Stellenbosch-7600, South Africa}

%\date{\today}

\date{\today}

%\maketitle

%\title{Quantum Geometric Phases as a New Window on Gravitational Waves}
%\author{Partha Nandi}
%\affiliation{National Institute for Theoretical and Computational Sciences (NITheCS), South Africa}
%\author{Frederik G. Scholtz}
%\affiliation{Department of Physics, University of Stellenbosch, Stellenbosch-7600, South Africa}
%\date{\today}

\begin{abstract}

We investigate how low-frequency gravitational waves (LFGWs), originating from distant astrophysical or cosmological sources, can induce purely quantum geometric phases in mesoscopic optomechanical systems—subtle imprints with no classical counterpart, beyond standard dynamical or Berry-type contributions that admit Hannay-angle analogues. These ultra-weak waves couple to the motion of a mechanical mirror and generate distinctive phase shifts in the system’s quantum state that cannot arise within any classical description. To access this effect, we propose a Ramsey-type interferometric protocol in which the photon-number states of a quantized optical mode become entangled with the mirror’s center-of-mass motion, enabling a direct readout of the LFGW-induced geometric phase. This framework establishes a distinctly quantum approach for probing low-frequency gravitational wave modes, offering an alternative to conventional detection strategies based on spacetime strain.

%We explore how low-frequency gravitational waves (LFGWs) can induce purely quantum geometric phases in mesoscopic optomechanical systems—subtle imprints with no classical analogue, beyond the standard dynamical and Berry-type contributions that admit Hannay-angle counterparts. Generated by distant astrophysical or cosmological sources, these ultra-weak waves couple to the motion of a mechanical mirror and imprint distinctive quantum phases that cannot arise in any classical description. To access this effect, we propose a Ramsey-type interferometric protocol in which the photon-number states of a quantized optical mode become entangled with the mirror's motion, enabling direct readout of the GWs-induced quantal geometric phase. This framework opens a conceptually new pathway for purely quantum mechanical probing of low-frequency gravitational wave modes through an interferometric-type setup, offering an alternative to conventional detection schemes.

\end{abstract}

\maketitle

\section{Introduction}

Although still a relatively young discipline, gravitational wave (GW) astronomy has already transformed fundamental physics. The breakthrough came with the first detection by LIGO~\cite{ligo2016}, which confirmed a key prediction of general relativity and inaugurated a new era of observational astrophysics. Gravitational waves—ripples in spacetime emitted by cataclysmic events such as black hole mergers—are typically detected by kilometer-scale laser interferometers that track minute displacements of suspended mirrors. Ground-based detectors like LIGO and Virgo are most sensitive to frequencies in the \(10\,\mathrm{Hz}\) to \(10^{3}\,\mathrm{Hz}\) range~\cite{PhysRevLett.120.141102}, but their sensitivity deteriorates rapidly below \(\sim10\,\mathrm{Hz}\), leaving the low-frequency GW spectrum largely unexplored.

This low-frequency regime is particularly important for cosmology and fundamental physics, encompassing signals from supermassive black hole binaries, cosmic strings, and relic gravitational waves from the early Universe~\cite{PhysRevD.103.103012}. While space-based observatories like LISA~\cite{PhysRevLett.116.231101} and pulsar timing arrays (PTAs)~\cite{Joshi2013PTA} are being developed to access this band, they rely on detecting displacements or timing residuals accumulated over large baselines. In the ultra-low-frequency domain, however, the induced mirror displacements become quasi-static and often fall below both classical and quantum noise floors. As a result, conventional detection schemes—whether laser-based or timing-based—face severe limitations.

For instance, in LIGO, a gravitational strain of \( h \sim 10^{-21} \) results in mirror displacements on the order of \( 10^{-18}\,\mathrm{m} \), while the quantum zero-point fluctuations of the mirrors---each weighing approximately \( 40\,\mathrm{kg} \)---are only about \( 10^{-20}\,\mathrm{m} \)~\cite{LIGOSuspension2012,Sigg1998GravitationalWaves,Nandi:2024jyf,Nandi:2024zxp}. However, at lower gravitational wave frequencies, two significant challenges arise. First, because the strain signal evolves more slowly, the induced displacement becomes correspondingly smaller---potentially approaching or falling below the detector's sensitivity threshold \cite{ThorneGravitationalRadiation}. This suppression is both technical, due to increased seismic and suspension noise at low frequencies, and fundamental, as the zero-point motion of the mirrors can exceed the curvature-induced displacement~\cite{RevModPhys.82.1155}.

Second—and more crucially—low-frequency gravitational waves induce slow, cyclic modulations in spacetime curvature, creating conditions that support adiabatic evolution. In such regimes, a quantum system can accumulate a global geometric phase that encodes the history of its trajectory through parameter space. This raises the compelling possibility that these \emph{geometric phases} could govern the system's response in a way that has no classical analogue---unlike the Hannay angle \cite{hannay1985}, which corresponds to classical adiabatic invariants. These phases are not captured by classical observables defined on phase space, such as position or momentum~\cite{Thompson:1993ts,Papini:1988qs}. Consequently, even in the absence of a measurable net displacement, the quantum state of the probe may retain a distinct and experimentally accessible imprint of the gravitational wave.

Accessing such imprints demands a detector capable of maintaining quantum coherence throughout the interaction. Macroscopic mirrors, like those used in LIGO, rapidly decohere and are therefore unsuitable for capturing coherent quantum effects such as geometric phase accumulation. In contrast, recent advances in optomechanics have opened a new experimental frontier: mesoscopic mechanical systems \cite{OConnell2010,Teufel2011,Chan2011}---with masses ranging from \(10^{-12}\) to \(10^{-9}\,\mathrm{kg}\)---can now be cooled to near their quantum ground states~\cite{PhysRevLett.91.130401,Adler2005}, rendering them suitable as quantum-coherent probes. In this intermediate mass range, the zero-point position uncertainty \(\Delta x_{\mathrm{zpf}} \sim 10^{-17}\,\mathrm{m}\) \cite{Miki:2023nce, Wan:2017thesis} becomes comparable to, or even larger than the gravitationally induced displacement. This convergence of quantum and gravitational scales signals not merely a technological advancement, but a qualitative shift in the principles of gravitational wave detection.

Recent proposals have explored the use of milligram-scale optomechanical systems in space-based interferometers to test gravitationally induced entanglement, demonstrating that decoherence can be suppressed via cryogenic shielding and feedback control~\cite{Matsumoto2025GIE}. These developments further underscore the experimental viability of mesoscopic quantum probes for exploring gravity in previously inaccessible regimes.

Against this backdrop, the concept of the \emph{geometric phase}---introduced by Berry in 1984~\cite{berry1984}---offers a powerful probe of spacetime curvature in quantum systems \cite{Papini:1988qs}. When a Hamiltonian depends on slowly varying parameters tracing a closed loop in parameter space, the system accumulates a total phase comprising a dynamical and a geometric component. The latter, the Berry phase, depends only on the path’s geometry and not on its traversal rate.

This phase often admits a classical analogue in the form of the Hannay angle~\cite{berry1985,hannay1985}, reflecting a shift in conjugate angle variables under adiabatic evolution. Aharonov and Anandan~\cite{aharonov1987} extended the concept beyond the adiabatic regime, while Simon~\cite{simon1983} recast it as a holonomy in a fiber bundle over parameter space, emphasizing its kinematic and geometric character. The geometric phase has since found profound connections in diverse areas, including canonical quantization of gravity~\cite{Surya:1997ej}, noncommutative spacetime~\cite{Chakraborty:2021uoc,Biswas:2019obt}, fractional statistics~\cite{arovas1984,haldane1985}, gauge anomalies~\cite{niemi1985,nelson1985,sonoda1986}, and the quantum Hall effect~\cite{thouless1982,semenoff1986}.

In prior work~\cite{nandi2023berry}, one of us demonstrated that a parametrically modulated quantum oscillator can acquire a Berry phase due to time-dependent weak gravitational curvature. While that analysis focused on isolated internal dynamics, it did not address how such phases might be detected or distinguished from dynamical contributions in a real experiment.

Building on that foundation, we develop a conceptually distinct and experimentally grounded framework for probing low-frequency gravitational wave modes through quantum phase evolution. Whereas earlier analyses focused exclusively on Berry phases arising from internal oscillator dynamics, our approach incorporates both external modulation and readout, allowing for direct access to phase information. Specifically, we consider a mechanical oscillator subject to a combination of low-frequency modes of classical GWs, constant radiation pressure, and time-dependent modulation of its trapping potential. In contrast to our previous treatment, we now explicitly include the effect of radiation pressure—an intrinsic and unavoidable feature of optomechanical systems—which induces coherent displacement of the oscillator’s wavepacket.

This extended structure gives rise to a new and previously unrecognized contribution: an \textit{Aharonov--Bohm--like phase}~\cite{aharonov1959,ehrenberg1949}, resulting from cyclic evolution in the Hamiltonian’s parameter space. This phase is qualitatively distinct from the familiar Berry phase (and its classical analog, the Hannay angle~\cite{hannay1985}): it originates from trajectory-dependent displacement rather than adiabatic squeezing, and encodes a global quantum phase that is invisible to classical dynamics. Although the classical system returns to its initial configuration without memory of the path, the quantum system retains a nontrivial phase that reflects the full history of its evolution. Together, the Berry and AB-like contributions reveal a uniquely quantum response to classical gravitational waves~\cite{Ruggiero:2024pzv}—providing a novel interferometric window into gravitational effects that elude conventional displacement-based measurements.

In this work, we propose a quantum interferometric scheme that measures the geometric phase shift acquired by a coherent quantum probe undergoing adiabatic evolution in curved spacetime. Unlike classical detectors such as LIGO or LISA, which register changes in path length due to spacetime strain, our approach is sensitive to the curvature-induced modulation of the quantum state itself. While conceptually distinct, it remains interferometric in nature and may be regarded as a quantum counterpart to LIGO---optimized for probing ultra-low-frequency gravitational waves (LFGWs) in a regime where classical displacement-based methods lose sensitivity.

Specifically, we demonstrate that:
\begin{itemize}
    \item Both the Berry and AB-like phases emerge from a unified Hamiltonian describing the mirror’s quantum evolution under LFGWs, radiation pressure, and trap modulation;
    \item The AB-like phase constitutes a path-dependent geometric phase with no classical analogue, arising from cyclic displacement in the Hamiltonian parameter space;
   % \item A Ramsey-type optomechanical interferometry protocol can isolate and extract this purely quantum phase shift, offering a concrete route to observe curvature-induced effects using mesoscopic systems.
\end{itemize}

%Our approach departs from conventional gravitational wave detection methods in two key respects. First, we uncover a previously unrecognized Aharonov--Bohm–like geometric phase that arises from coherent displacement in a time-dependent Hamiltonian, induced by low-frequency gravitational waves (LFGWs), even in the absence of any net displacement or proper-time difference. Second, we propose a realistic Ramsey-type interferometric protocol capable of isolating this purely quantum phase using mesoscopic mechanical systems subject to radiation pressure. Together, these features enable direct access to quantum signatures of gravity in a parameter regime where standard displacement-based detectors become ineffective.

Both phases are derived explicitly from the time-dependent Hamiltonian governing the quantum evolution of the mirror. While the Berry phase admits a classical analog via the Hannay angle, the AB-like phase is uniquely quantum, reflecting a path-dependent memory encoded in the wavefunction. This phase depends on the full trajectory in parameter space and is invisible to classical equations of motion.

It is important to emphasize that the term ``Aharonov--Bohm--like phase'' in our context does \emph{not} refer to the conventional electromagnetic Aharonov--Bohm (AB) effect in physical space---where the electromagnetic potentials acquire observable significance even in regions devoid of electric or magnetic fields. Rather, it denotes a \emph{structural analogue} of the AB effect, arising in the \emph{parameter space} of the system's Hamiltonian. In the original AB effect, a charged particle accumulates a quantum phase due to the presence of a vector potential, despite following a trajectory through a field-free region. The classical dynamics remain unchanged, but the quantum wavefunction retains a global, gauge-invariant phase shift that depends only on the enclosed magnetic flux---a \emph{topological} feature.

In our system, a similar phenomenon occurs. gravitional waves with low frequency and radiation pressure modulate the Hamiltonian such that the quantum state traces a closed loop in parameter space. Although the classical trajectory of the mirror returns unchanged, the quantum system acquires a global geometric phase that reflects the full history of the evolution. This phase acts as a quantum memory of the path traversed, while the classical system “forgets” it. Though not topological, this phase is geometric and global, arising from a gauge-like structure that is only accessible to quantum dynamics.

To measure these phases, we propose a Ramsey-type optomechanical interferometric scheme in which a suspended mirror is entangled with a quantized optical mode or qubit~\cite{Ramsey1950}. This marks a shift from a classical to a quantized treatment of radiation pressure: the single-photon state both induces a discrete force on the mirror and serves as a phase-sensitive readout. Such quantized fields are natural in cavity-based Ramsey protocols, enabling detection at the quantum limit. Under symmetric conditions—e.g., identical squeezing in both arms—the Berry phase cancels in the differential signal, isolating the purely quantum AB-like contribution. This allows direct access to curvature-induced geometric phases with no classical analog.

The remainder of this paper is structured as follows. Section II reviews linearized gravitational waves and their coupling to matter. In Section III, we formulate the quantum dynamics of a mesoscopic optomechanical system under curvature and radiation pressure. Section IV derives the resulting Berry and AB-like geometric phases, and Section V outlines a Ramsey-type interferometric scheme for their detection. We conclude in Section VI with a discussion of implications and outlook.

\section{Linearized Gravitational Waves and Effective Detector Dynamics}

In the weak-field limit of general relativity \cite{maggiore2008}, gravitational waves (GWs) are described as small perturbations \( h_{\mu\nu} \) to the flat Minkowski background \( \eta_{\mu\nu} \):
\begin{equation}
g_{\mu\nu} = \eta_{\mu\nu} + h_{\mu\nu}, \quad |h_{\mu\nu}| \ll 1.
\end{equation}
To second order, the Einstein-Hilbert action becomes:
\begin{equation}
\begin{aligned}
S_{EH} = \frac{1}{64 \pi G} \int d^4x \Big( 
& h_{\mu\nu} \Box h^{\mu\nu} + 2 h^{\mu\nu} \partial_\mu \partial_\nu h \\
& - h \Box h - 2 h_{\mu\nu} \partial_\rho \partial^\mu h^{\nu\rho} \Big).
\end{aligned}
\end{equation}

Under infinitesimal coordinate transformations generated by a vector field \( \zeta^\mu \), the metric perturbation transforms as
\begin{equation}
h_{\mu\nu} \rightarrow h_{\mu\nu} + \partial_\mu \zeta_\nu + \partial_\nu \zeta_\mu.
\end{equation}

To eliminate gauge redundancy and simplify the field equations, one imposes the \emph{harmonic gauge} (or de Donder gauge), defined by the condition:
\begin{equation}
\partial^\mu \bar{h}_{\mu\nu} = 0, \quad \text{where} \quad \bar{h}_{\mu\nu} = h_{\mu\nu} - \frac{1}{2} \eta_{\mu\nu} h,
\end{equation}
with \( h = \eta^{\mu\nu} h_{\mu\nu} \) being the trace of the perturbation. Under this gauge, the linearized Einstein field equations in vacuum reduce to simple wave equations:
\begin{equation}
\Box \bar{h}_{\mu\nu} = 0.
\end{equation}

To further isolate the physical degrees of freedom, it is common to impose the \emph{transverse-traceless (TT) gauge}. This is defined with respect to a static observer with four-velocity \( u^\mu = (1, 0, 0, 0) \), and imposes the following conditions:
\begin{equation}
h^{TT}_{0\mu} = 0, \quad \partial^i h^{TT}_{ij} = 0, \quad h^{TT~\mu}_{\ \mu} = 0.
\end{equation}
In this gauge, the remaining components \( h^{TT}_{ij} \) satisfy the vacuum wave equation:
\begin{equation}
\Box h^{TT}_{ij}(t, \vec{x}) = 0, \quad i,j = 1,2,3,
\end{equation}
and describe freely propagating gravitational waves with two polarization modes.

To understand how gravitational waves affect matter, we consider a congruence of timelike geodesics \( x^{\mu}(\tau) \) centered around the reference worldline of a freely falling detector. The analysis is performed in the \emph{proper detector frame}, a locally inertial coordinate system constructed along the detector’s trajectory \cite{Antoniou2016PropagationGW}. In this frame, the metric is locally Minkowskian at the origin, but tidal effects due to spacetime curvature manifest at leading order in spatial separation. The four-velocity of the detector is \( u^\mu=\frac{dx^{\mu}}{d\tau} = (1, 0, 0, 0) \), and we take the separation vector \( \xi^\mu \) to be spacelike and orthogonal to the reference geodesic, satisfying \( \xi^\mu u_\mu = 0 \) \cite{maggiore2008}. This ensures that \( \xi^\mu \) represents the physical spatial displacement between nearby geodesics, as measured by a local observer attached to the detector.

The relative acceleration between neighboring geodesics is governed by the geodesic deviation equation \cite{MTW:1973}:
\begin{equation}
\frac{D^2 \xi^\mu}{D \tau^2} = - R^\mu_{\ \nu\rho\sigma} \, \xi^\rho \, u^\nu u^\sigma,
\end{equation}
where \( R^\mu_{\ \nu\rho\sigma} \) is the Riemann curvature tensor evaluated along the detector’s worldline. In linearized gravity, it takes the form:
\begin{equation}
R^\mu_{\ \rho\sigma\nu} = \frac{1}{2} \eta^{\mu\lambda} \left( \partial_\sigma \partial_\rho h_{\nu\lambda} - \partial_\sigma \partial_\lambda h_{\nu\rho} - \partial_\nu \partial_\rho h_{\sigma\lambda} + \partial_\nu \partial_\lambda h_{\sigma\rho} \right).
\end{equation}

To proceed, we adopt the \emph{long-wavelength approximation}, which is valid when the gravitational wavelength is much larger than the spatial extent of the detector. In this limit, the metric perturbation becomes effectively time-dependent:
\begin{equation}
h^{TT}_{ij}(t, \vec{x}) \approx h^{TT}_{ij}(t).
\end{equation}
Substituting into the geodesic deviation equation and working in the proper detector frame, the spatial components get simplify to:
\begin{equation}
\frac{d^2 \xi^i}{dt^2} = \frac{1}{2} \ddot{h}^{TT}_{ij}(t) \, \xi^j.
\end{equation}

Assuming the gravitational wave propagates along the \( x_3 \) (or \( z \)) direction, this approximation captures the tidal effects of the wave solely through the time derivatives of the metric perturbation. In the transverse-traceless (TT) gauge, the general form of a plane wave solution \cite{MTW:1973} is given by
\begin{equation}
h^{TT}_{ij}(t) = \chi(t) \left( \epsilon_\times \, \sigma^1_{ij} + \epsilon_+ \, \sigma^3_{ij} \right),
\end{equation}
where \( \chi(t) \) denotes the wave amplitude, typically chosen for linearly polarized waves with a time-dependent modulation such as \( \cos(\omega_g t) \), with \(\omega_g\) being the frequency of the gravitational waves. The parameters \( \epsilon_\times \) and \( \epsilon_+ \) represent the polarization amplitudes, while \( \sigma^1_{ij} \) and \( \sigma^3_{ij} \) are the standard basis tensors corresponding to the two linear polarization states.

%Assuming that the gravitational wave propagates along the \( x_3 \) (or \( z \)) direction, this approximation captures the wave’s tidal effects solely through the time derivatives of the perturbation. In the transverse-traceless (TT) gauge, the general form of a plane wave \cite{MTW:1973} solution is:
%\begin{equation}
%h^{TT}_{ij}(t) = 2 \chi(t) \left( \epsilon_\times \, \sigma^1_{ij} + \epsilon_+ \, \sigma^3_{ij} \right),
%\end{equation}
%Here, \( \chi(t) \) denotes the amplitude which typically chosen for linearly polarized waves as with time dpendnt modulation  $cos(\omega_{g}t)$ of the gravitational wave, while \( \epsilon_\times \) and \( \epsilon_+ \) represent the polarization amplitudes. The tensors \( \sigma^1_{ij} \) and \( \sigma^3_{ij} \) are the standard basis tensors corresponding to the two linear polarization states.

\subsection{Including External Forces: Effective Detector Model}

We consider a mesoscopic optomechanical mirror system~\cite{PhysRevA.87.043832}, modeled via its center-of-mass (COM) motion as an effective point particle constrained to move along the \( x \)-axis, and subject to two external forces: a classical electromagnetic (EM) field exerting constant radiation pressure, and a background gravitational wave.

Working in Fermi normal coordinates centered at a fixed laboratory reference point (e.g., the optical trap center), the mirror's motion is subject to both curvature-induced tidal accelerations and externally controlled optical forces. The effective classical equation of motion for the center-of-mass displacement \( \xi(t) \) is
\begin{equation}
m \ddot{\xi}_{1}(t) + m \omega_0^2(t) \, \xi_{1}(t) = F_{\mathrm{rad}} + \frac{m}{2} \ddot{h}^{T}_{11}(t) \, \xi_{1}(t),
\end{equation}
where \( m \) is the mirror mass, \( \omega_0(t) \) is a (possibly time-dependent) trapping frequency controlled via optical means, \( F_{\mathrm{rad}} \) is the radiation pressure force, and \( h^{T}_{11}(t) \) is the gravitational wave strain in the transverse-traceless gauge. 

%This classical formulation serves as the starting point for our subsequent quantum analysis, where the mirror is treated as a coherent mesoscopic probe capable of registering curvature-induced quantum phases.

%We consider a suspended mirror at the end of an interferometer arm, treated as a point particle constrained to move along the \( x \)-axis. The mirror interacts with two external forces: a background gravitational wave and a classical electromagnetic (EM) field that exerts constant radiation pressure.

%Working in Fermi normal coordinates centered at the beam splitter, the mirror's motion is subject to both the curvature-induced tidal acceleration and the externally applied optical force. The effective classical equation of motion for the mirror's center-of-mass displacement \( \xi(t) \) is
%\begin{equation}
%m \ddot{\xi}_{1}(t) + m \omega_0^2(t) \, \xi_{1}(t) = F_{\mathrm{rad}} + \frac{m}{2} \ddot{h}^{T}_{11}(t) \, \xi_{1}(t),
%\end{equation}
%where \( m \) is the mirror mass, \( \omega_0(t) \) is a possibly time-dependent trapping frequency (from optical springs or suspensions), \( F_{\mathrm{rad}} \) is the constant radiation pressure force, and \( h^{T}_{11}(t) \) is the GW strain component along the interferometer arm.

It is important to note that the time dependence of \( \omega_0(t) \) is externally controlled (e.g., via modulation of the optical cavity or trapping potential) and is independent of the gravitational wave. In contrast, the GW enters through the term \( \ddot{h}^{T}_{11}(t) \), which modifies the tidal acceleration. This distinction ensures that the GW does not alter the potential, but rather induces an additional driving term.

The equation of motion can be derived from an effective Lagrangian~\cite{nandi2023berry,Nandi:2024zxp}:
\begin{equation}
L_D = \frac{1}{2} m \dot{\xi}_{1}^2 - m \dot{h}^{T}_{11}(t) \dot{\xi}_{1} \xi_{1} - V_{\mathrm{ext}}(\xi_{1}, t),
\end{equation}
where \( V_{\mathrm{ext}}(\xi_{1}, t) \) includes the trap potential and radiation pressure. The corresponding Hamiltonian reads
\begin{equation}
H_D = \frac{p_{1}^2}{2m} + \frac{1}{2} \dot{h}^{T}_{11}(t) \, \xi_{1} \, p_{1} + V_{\mathrm{ext}}(\xi_{1}, t).
\end{equation}

A closely related form of this Hamiltonian was recently studied in Ref.~\cite{Parikh:2021} to investigate the quantum nature of gravity within a two-particle detector model. An earlier variant, originally introduced in Ref.~\cite{Speliotopoulos:1995}, has also been employed in the context of noncommutative quantum mechanics for different purposes (see, e.g., Ref.~\cite{Saha:2018}). In our work, this Hamiltonian serves as the foundation for a fully quantum treatment of the mirror’s motion and its coupling to the gravitational wave background, which we develop in the following section.

\section{Quantum Dynamics of the Mirror and Gravitationally Induced Geometric Phases}

In conventional gravitational wave detectors such as LIGO, the end mirrors are suspended as pendulums with masses on the order of \( 40\,\mathrm{kg} \) and fundamental suspension frequencies around \( 0.74\,\mathrm{Hz} \)~\cite{LIGOSuspension2012}. The gravitational-wave-induced displacement (\( \sim 10^{-18}\,\mathrm{m} \)) is roughly an order of magnitude larger than the quantum zero-point fluctuation of the mirror, which remains unresolvable due to the mirror's high thermal occupation at room temperature. As a result, LIGO does not probe the quantum motion of the mirror, and classical models suffice for describing its dynamics. Therefore, intrinsically quantum effects such as  Aharonov–Bohm–like phenomena, which rely on coherent quantum evolution in presence of GWs, are unobservable in such macroscopic setups.

In contrast, mesoscopic optomechanical platforms employ mechanical oscillators with effective masses in the range \( 10^{-12} \) to \( 10^{-9}\,\mathrm{kg} \)~\cite{vonLupke2022, Aspelmeyer2014, Vitali2007}. In such systems, the quantum zero-point position uncertainty is given by
\begin{equation}
    \Delta x_{\mathrm{zpf}} = \sqrt{\frac{\hbar}{2 m \omega_0}},
\end{equation}
where \( \omega_0 \) denotes the trap frequency of the oscillator. For representative devices with frequencies in the range \( \omega_0 \sim 10^5 \text{–} 10^6\,\mathrm{rad/s} \) and effective masses \( m \sim 10^{-11} \text{–} 10^{-10}\,\mathrm{kg} \), the zero-point fluctuations reach \( \Delta x_{\mathrm{zpf}} \sim 10^{-17}\,\mathrm{m} \), as demonstrated in recent experiments~\cite{Teufel2011, Chan2011, vonLupke2022}. Crucially, such systems can be cooled to near their quantum ground state, with thermal occupation numbers \( \langle n \rangle < 1 \) achieved through cryogenic and sideband cooling~\cite{Teufel2011, Chan2011, Delic2020}. In this regime, thermal noise becomes negligible, and the oscillator exhibits coherent quantum dynamics. In such regimes, it is the center-of-mass (COM) degree of freedom of the mechanical oscillator that undergoes coherent quantum evolution. One might argue that for any realistic mesoscopic object, internal vibrational or thermal modes could couple to the COM and cause decoherence, washing out any interferometric phase signature. However, our system is based on the geodesic deviation equation, which governs the COM motion relative to a reference geodesic and is valid to leading order in the absence of internal gradients. Moreover, for translation-invariant potentials that depend only on displacement from a fixed reference point (as is the case for harmonic traps or optical springs), the COM mode remains decoupled from internal degrees of freedom. This justifies our focus on the COM coordinate as the primary probe of gravitationally induced geometric phases. This makes mesoscopic optomechanical systems ideal candidates for probing gravitationally induced quantum responses beyond classical displacement. In such regimes, the quantum nature of the mechanical probe becomes essential, as it enables sensitivity to phase shifts that would be invisible to classical displacement-based measurements. This is particularly important when the gravitational wave frequency \( \omega_{\mathrm{gw}} \) lies well below the oscillator frequency \( \omega_0 \), ensuring adiabatic response and enabling coherent phase accumulation.

%In contrast, mesoscopic optomechanical platforms utilize mechanical oscillators with masses in the range \( 10^{-12} \) to \( 10^{-9}\,\mathrm{kg} \), for which the quantum zero-point position uncertainty,
%\begin{equation}
%\Delta x_{\mathrm{zpf}} = \sqrt{\frac{\hbar}{2m\omega}},
%\end{equation}
%can reach \( \sim 10^{-17}\,\mathrm{m} \)—comparable to or even exceeding displacements induced by gravitational waves. In such regimes, a quantum mechanical treatment becomes essential to capture GWs-sensitive phase dynamics.

Motivated by advances in cavity optomechanics and recent proposals for gravitational-wave-induced quantum effects~\cite{Matsumura}, we model the mirror as a quantum harmonic oscillator governed by a time-dependent Hamiltonian. In particular, to enable adiabatic geometric phase accumulation and selectively enhance coupling to gravitational wave modes, we promote the trap frequency to a slowly varying, periodic function of time:
\[
\omega_0 \rightarrow \omega_0(t) = \omega_0 \kappa(t),
\]
where \( \kappa(t) \) a dimensionless, smooth, and externally tunable periodic modulation function. Such modulation can be engineered via optical control \cite{PhysRevLett.66.527} , and plays a role analogous to frequency tuning in a radio receiver—allowing the mechanical system to resonate with specific low-frequency gravitational wave components.

The resulting time-dependent Hamiltonian takes the form:
\begin{equation}
\hat{H}_D(t) = \frac{1}{2} \left( Z \hat{p}_1^2 + X(t) \hat{\xi}_1^2 + Y_g(t)(\hat{\xi}_1 \hat{p}_1 + \hat{p}_1 \hat{\xi}_1) \right) - F_{\mathrm{rad}} \hat{\xi}_1,
\end{equation}
where \( Z = 1/m_0 \), \( X(t) = m_0 \omega_0^2(t) = m_0 \omega_0^2 \kappa^2(t) \) encodes the modulated trap stiffness, and \( Y_g(t) = \dot{\chi}(t)\epsilon_+ \) represents the tidal coupling to gravitational waves through the strain amplitude \( \chi(t) \). The final term accounts for a constant radiation pressure force \( F_{\mathrm{rad}} \). In this framework, the time-dependent scaling \( \kappa(t) \) acts as both a geometric deformation parameter and a dynamical control channel for gravitational signal enhancement.

%Motivated by advances in cavity optomechanics and proposals for GWs-induced quantum effects~\cite{Matsumura}, we model the mirror as a quantum harmonic oscillator governed by a time-dependent Hamiltonian:
%\begin{equation}
%\hat{H}_D(t) = \frac{1}{2} \left( Z \hat{p}_1^2 + X(t) \hat{\xi}_1^2 + Y_g(t)(\hat{\xi}_1 \hat{p}_1 + \hat{p}_1 \hat{\xi}_1) \right) - F_{\mathrm{rad}} \hat{\xi}_1,
%\end{equation}
%where \( Z = 1/m_0 \), \( X(t) = m_0 \omega_0^2(t) \) encodes the time-dependent trap frequency, and \( Y_g(t) = \dot{\chi}(t)\epsilon_+ \) accounts for gravitational waves. The system also experiences a constant radiation pressure force \( F_{\mathrm{rad}} \).

The time-dependent Hamiltonian can be mapped to a displaced harmonic oscillator via a unitary transformation:
\begin{equation}
\hat{H}_D(t) = \hat{U}(t)\, \hat{H}_{\mathrm{ho}}(t)\, \hat{U}^\dagger(t),
\end{equation}
where the effective harmonic oscillator Hamiltonian is given by:
\begin{equation}
\hat{H}_{\mathrm{ho}}(t) = \frac{1}{2} \left( Z \hat{p}_1^2 + \frac{\omega^2(t)}{Z} \hat{\xi}_1^2 \right) - \frac{F_{\mathrm{rad}}^2 Z}{2 \omega^2(t)}.
\end{equation}

The time-dependent frequency is defined as:
\begin{equation}
\omega^2(t) = X(t) Z - Y_g^2(t) > 0 \quad \text{for all } t,
\label{t}
\end{equation}
and the full unitary operator \( \hat{U}(t) \) is composed of two successive transformations:
\begin{equation}
\hat{U}(t) = \hat{U}_2(t) \hat{U}_1(t),
\end{equation}
with
\begin{equation}
\hat{U}_2(t) = \exp\left( -\frac{i Y_g(t)}{2\hbar Z} \hat{\xi}_1^2 \right), \quad
\hat{U}_1(t) = \exp\left( i \frac{F_{\mathrm{rad}} Z}{\omega^2(t) \hbar} \hat{p}_1 \right).
\end{equation}

Here, \( \hat{U}_1(t) \) generates a coherent displacement in position space, while \( \hat{U}_2(t) \) implements a squeezing-like transformation arising from the parametric time dependence in the Hamiltonian.

The instantaneous eigenstates of the full Hamiltonian are given by
\begin{equation}
|n(t)\rangle_D = \hat{U}(t) |n(t)\rangle_{\mathrm{ho}}, 
\quad 
E_n(t) = \hbar \omega(t)\left(n + \frac{1}{2}\right) - \frac{F_{\mathrm{rad}}^2 Z^2}{2 \omega^2(t)}.
\label{gh}
\end{equation}

A key feature of our setup is the synchronization of the detector's frequency modulation with the frequency of the gravitational wave background. This condition is not merely convenient—it is essential. The emergence of a nontrivial Berry phase requires that the system evolve adiabatically along a closed loop in parameter space. This necessitates that all time-dependent parameters—such as the oscillator frequency and the curvature-induced coupling—share the same periodicity. In our framework, this requirement is naturally satisfied, as the gravitational wave modulates both curvature and trap frequency on the same time scale. This ensures the cyclicity of the Hamiltonian and satisfies the conditions of the adiabatic theorem, enabling the accumulation of a nontrivial geometric phase.

Under adiabatic evolution over a period \( T \), the quantum state acquires both  dynamical and geometric contribution:
\begin{equation}
|\Psi(T)\rangle = e^{-\frac{i}{\hbar} \int_0^T E_n(t)\,dt} \, e^{i\Phi_G} \, |n(T)\rangle_D,
\end{equation}
where the geometric phase is given by
\begin{align}
\Phi_G &= i \int_0^T \langle n(t)| \frac{d}{dt} |n(t)\rangle_D dt \nonumber\\
&= i\int_0^T \langle n(t)| \hat{U}^\dagger(t) \frac{d}{dt} \hat{U}(t) |n(t)\rangle_{\mathrm{ho}} dt.
\label{hd}
\end{align}

Using the following time derivatives of the unitary operators defined in Eq.~(\ref{hd}), we find:
\begin{equation}
\begin{aligned}
\dot{\hat{U}}_2 &= -\frac{i}{2\hbar Z} \, \dot{Y}_g(t)\, \hat{\xi}_1^2 \, \hat{U}_2, \\
\dot{\hat{U}}_1 &= i\, \frac{d}{dt} \left( \frac{F_{\mathrm{rad}} Z}{\omega^2(t)} \right)
\frac{1}{\hbar} \hat{p}_1 \, \hat{U}_1,
\end{aligned}
\end{equation}

the geometric phase accumulated over the evolution can be explicitly evaluated as:
\begin{equation}
\begin{aligned}
\Phi_G = -\int_0^T dt \bigg[
&\left(n + \frac{1}{2} \right) \frac{Z}{2 \omega(t)} 
\frac{d}{dt} \left( \frac{Y_g(t)}{Z} \right) \\
&+ \frac{1}{2\hbar} a^2(t) 
\frac{d}{dt} \left( \frac{Y_g(t)}{Z} \right)
\bigg],
\end{aligned}
\end{equation}
where \( a(t) = \frac{F_{\mathrm{rad}} Z}{\omega^2(t)} \) encodes the coherent displacement of the wavepacket, induced by external influences such as radiation pressure and gravitational waves.

In deriving this expression, we use the expectation value of the squared position operator in a harmonic oscillator eigenstate:
\begin{equation}
\langle n(t)|\hat{\xi}_1^2|n(t)\rangle_{\text{ho}} = \frac{\hbar}{2m_0\omega(t)}(2n + 1),
\end{equation}
and that the adiabatic derivative term vanishes:
\begin{equation}
\langle n(t)|i\frac{\partial}{\partial t}|n(t)\rangle_{\text{ho}} = 0,
\end{equation}
since the eigenfunctions of the harmonic oscillator are real-valued and therefore do not contribute to the adiabatic geometric phase.

For a closed loop \( \mathcal{C} \) in parameter space \( \vec{R}(t) = (\omega_0(t), Y_g(t)) \), the geometric phase acquired by the system takes the standard gauge-invariant form:
\begin{equation}
\Phi_G[\mathcal{C}] = \oint_{\mathcal{C}} d\vec{R} \cdot \vec{A}(\vec{R}),
\end{equation}
where \( \vec{A}(\vec{R}) \) is the adiabatic connection given by:

\begin{equation}
\vec{A}(\vec{R}) = \vec{A}^{(n)}_B(\vec{R}) + \vec{A}_{AB}(\vec{R}),
\end{equation}
with
\begin{equation}
\begin{aligned}
\vec{A}^{(n)}_B(\vec{R}) &= -\left(n + \frac{1}{2} \right) \frac{Z}{2\omega} 
\nabla_{\vec{R}} \left( \frac{Y_g}{Z} \right), \\
\vec{A}_{AB}(\vec{R}) &= -\frac{1}{2\hbar} a^2 
\nabla_{\vec{R}} \left( \frac{Y_g}{Z} \right).
\end{aligned}
\end{equation}

The first term \( \vec{A}^{(n)}_B \) corresponds to the standard Berry connection, explicitly dependent on the quantum number \( n \), and has a classical analog in the form of a finite Hannay angle \cite{PhysRevD.37.1709}. The second term \( \vec{A}_{AB} \), however, lacks any classical analogue: it stems from the coherent displacement of the quantum wavepacket and leads to an Aharonov--Bohm--like phase. This term reflects the intrinsic gauge structure of the Hamiltonian’s parameter space and contributes a purely quantum correction to the geometric phase, detectable through interferometric measurements, despite leaving classical trajectories unaffected.

\subsection{Explicit Evaluation of Berry and AB-like Phases}

To evaluate the Berry and AB-like phases analytically, we consider a periodic modulation of the trap frequency and the curvature induced by low-frequency gravitational wave (GW) modes over a single cycle \( T = 2\pi / \omega_g \), where \( \omega_g \) is the GW frequency. Specifically, we adopt a time-dependent dimensionless scaling function  of the form
\begin{equation}
\kappa(t) = \sqrt{ \left(1 + \frac{\Omega}{\omega_0} \cos(\omega_g t) \right)^2 
+ \left( \frac{\nu_0}{\omega_0} \right)^2 \sin^2(\omega_g t) },
\label{p}
\end{equation}
where \( \Omega \), \( \omega_0 \), and \( \nu_0 \) are regarded as time-independent and externally adjustable parameters.

This choice ensures that the total frequency (\ref{t}) remains positive and periodic, and matches the characteristic timescales of the detector with the external gravitational wave frequency. To capture the GW interaction explicitly, we introduce the gravitational wave interaction coupling parameter, denoted as
\begin{equation}
Y_g(t) = \dot{\chi}(t) \epsilon_+ = \omega_g \chi_0 \epsilon_+ \sin(\omega_g t),
\label{b}
\end{equation}
which guarantees the synchronization of the time periods of the detector’s frequency parameters with the low-frequency gravitational wave’s frequency. Such synchronization is particularly important, as discussed below Eq.~(\ref{gh}).

By substituting into the Berry-phase expression and making the change of variable \( z = e^{i \omega_g t} \), we can rewrite the Berry phase as a contour integral along the unit circle \( |z| = 1 \):

\begin{align}
\Phi^{(n)}_B &= \oint_{\mathcal{C}} d\vec{R} \cdot \vec{A}_B(\vec{R}) \notag \\
&= -\frac{i \,\omega_g \chi_0 \,\epsilon_+}{\Omega} \left(n+\frac{1}{2}\right)
\oint_{|z|=1} \frac{z^2 + 1}{z(z^2 + 2 a_0 z + 1)} \, dz, \notag \\
&\quad \text{with} \quad a_0 = \frac{\omega_0}{\Omega}.
\end{align}

Using the residue theorem to evaluate the contour integral, we obtain

\begin{align}
\Phi^{(n)}_B &= \left(n + \frac{1}{2}\right) 
\frac{2\pi \,\omega_g \chi_0 \,\epsilon_+}{\Omega} 
\left(1 - \frac{1}{\sqrt{1 - \varepsilon}} \right), \notag \\
&\quad \varepsilon = \frac{1}{a_0^2}, \quad 0 < \varepsilon < 1.
\end{align}

Similarly, for the Aharonov--Bohm--like geometric phase under the influence of low-frequency gravitational waves, the corresponding contour expression is:
\begin{multline}
\Phi_{\mathrm{AB}} = \oint_{\mathcal{C}} d\vec{R} \cdot \vec{A}_{\mathrm{AB}}(\vec{R}) \\
= -\frac{i\omega_g \chi_0 \,\epsilon_+}{\Omega^2}\,\frac{F_{\mathrm{rad}}^2 Z}{2\hbar}\,
\oint_{|z| = 1} \frac{\,z^2 + 1\,}{\,z^2 (z^2 + 2 a_{0} z + 1)^2\,}\, dz\,
\end{multline}
which, upon computing the residues, simplifies to:

\begin{equation}
\Phi_{\mathrm{AB}} = \frac{\pi \omega_g \chi_0 \epsilon_+ F_{\mathrm{rad}}^2 Z}{\hbar \Omega^2}
\left[ -4a_0 - \mathcal{F}(a_0) \right],
\end{equation}
where the function \( \mathcal{F}(a_0) \) encodes the GWs-dependent correction:
\begin{equation}
\mathcal{F}(a_0) = 
\frac{(a_0 - \sqrt{a_0^2 - 1})(1 + 2a_0^2) + 2a_0(a_0^2 - 1)}
{2(a_0^2 - 1)^{3/2}(a_0 - \sqrt{a_0^2 - 1})^3}.
\end{equation}

\vspace{0.5em}
In the small-\( \varepsilon \) limit, with \( \varepsilon = 1/a_0^2 \ll 1 \), this simplifies to
\begin{equation}
\Phi_{\mathrm{AB}} \approx 
- \frac{4\pi \omega_g \chi_0 \epsilon_+ F_{\mathrm{rad}}^2 Z}{\hbar \Omega^2 \sqrt{\varepsilon}} + \mathcal{O}(\varepsilon).
\end{equation}

Both phase shifts scale linearly with the gravitational wave amplitude \( \chi_0 \epsilon_+ \), but arises from distinct physical mechanisms:
\begin{itemize}
    \item The \textbf{Berry phase} originates from adiabatic squeezing induced by the LFGWs.
    \item The \textbf{AB-like phase} captures the cumulative effect of radiation pressure displacing the oscillator in the background of LFGWs modes.
\end{itemize}

Importantly, this AB-like phase is \textit{purely quantum geometric phases}, undetectable via classical displacement-based methods. Their dependence on tunable parameters such as \( \Omega \), \( \omega_0 \), and \( \omega_g \) offers concrete experimental handles.

Having derived these GW-sensitive phases, we now present an interferometric protocol to extract them experimentally.

\section{Interferometric Detection of Gravitationally Induced Phases}

To measure the gravitational wave–induced quantum phase shifts discussed earlier, we propose a Ramsey-type optomechanical interferometric scheme involving a mesoscopic mechanical mirror entangled with a quantized optical mode or auxiliary qubit~\cite{Ramsey1950}. While earlier sections treated radiation pressure as a classical force, we now adopt a fully quantized description of the optical field. This transition is natural in the Ramsey framework, where a single photon not only imparts discrete momentum to the mirror but also serves as the internal degree of freedom enabling interferometric readout. The resulting phase-sensitive measurement scheme captures purely quantum features of gravitational interaction that would otherwise be inaccessible. The procedure unfolds through the following steps.

%\lipsum[1]  % filler text

\begin{figure*}[t]
\centering
\begin{tikzpicture}[node distance=0.9cm and 1.2cm, every node/.style={font=\footnotesize}, >=Stealth]

% Nodes
\node (init) [draw, rectangle, rounded corners, fill=blue!10] {Initial state: $|\psi(0)\rangle_m \otimes |0\rangle$};

\node (pi1) [draw, rectangle, rounded corners, below=of init, fill=green!10, align=center] {First Weak Coherent Pulse: \\ $(|0\rangle + |1\rangle)/\sqrt{2}$};

\node (split) [below=0.3cm of pi1] {};
\node (B) [draw, rectangle, rounded corners, left=of split, fill=orange!15, align=center] 
{Branch B:\\ $\hat{H}_B(t)$\\ $F_{\text{rad}} = 0$};

%\node (B) [draw, rectangle, rounded corners, left=of split, fill=orange!15] {{Branch B: \\ $\hat{H}_B(t)$, \\ $F_{\text{rad}} = 0$}};

\node (A) [draw, rectangle, rounded corners, right=of split, fill=orange!15, align=center] 
{Branch A:\\ $\hat{H}_A(t)$\\ $F_{\text{rad}} = \hbar g$};

%\node (A) [draw, rectangle, rounded corners, right=of split, fill=orange!15] {Branch A: \\ $\hat{H}_A(t)$, \\ $F_{\text{rad}} = \hbar g$};

\node (merge) [below=1.1cm of split] {};

\node (pi2) [draw, rectangle, rounded corners, below=of merge, fill=green!10] {Second Weak Coherent Pulse};
\node (meas) [draw, rectangle, rounded corners, below=of pi2, fill=red!10, align=center] 
{Readout:\\ $P_0 = \frac{1}{2}[1 + \cos(\Delta \Phi)]$};

%\node (meas) [draw, rectangle, rounded corners, below=of pi2, fill=red!10] {Readout: \\ $P_0 = \frac{1}{2}[1 + \cos(\Delta \Phi)]$};

% Arrows
\draw[->] (init) -- (pi1);
\draw[->] (pi1) -- (split);
\draw[->] (split) -- (A);
\draw[->] (split) -- (B);
\draw[->] (A) -- (merge);
\draw[->] (B) -- (merge);
\draw[->] (merge) -- (pi2);
\draw[->] (pi2) -- (meas);

\end{tikzpicture}
\caption{Ramsey interferometry protocol to detect AB-like and dynamical phases induced by gravitational waves and radiation pressure.}
\label{fig:ramsey_diagram}
\end{figure*}
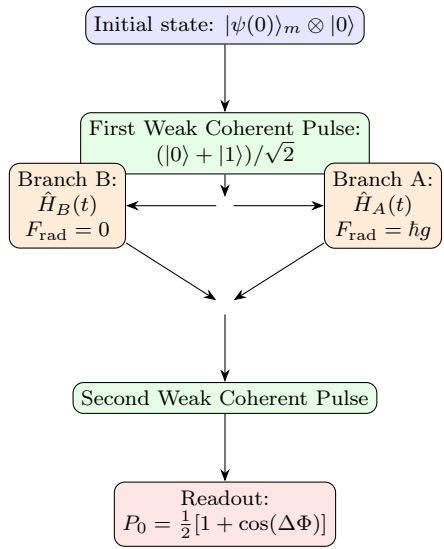

%\lipsum[2]  % filler text

\subsection{Step 1: Initial State Preparation}

We consider a hybrid quantum system consisting of a mesoscopic mechanical mirror coupled to an effective two-level system (TLS), implemented via the interaction with a weak radiation pulse. In the ideal case, this pulse is prepared in the single-photon Fock state \(|1\rangle\). Initially, the combined system is prepared in a product state:
\begin{equation}
|\Psi(0)\rangle = |\psi(0)\rangle_m \otimes |0\rangle,
\end{equation}
where \(|\psi(0)\rangle_m\) denotes the initial motional state of the mirror, and \(|0\rangle\) represents the vacuum state of the radiation field—that is, the ground state of the effective TLS.

\subsection{Step 2: State Preparation via First Weak Coherent Pulse}

While our theoretical model invokes a Hadamard-like transformation to prepare a balanced superposition of vacuum and single-photon components,
\begin{equation}
    |0\rangle \rightarrow \frac{1}{\sqrt{2}} (|0\rangle + |1\rangle),
\end{equation}
resulting in the entangled initial state
\begin{equation}
    |\Psi_0(0)\rangle = \frac{1}{\sqrt{2}} |\psi(0)\rangle_m \otimes (|0\rangle + |1\rangle),
\end{equation}
this operation is to be understood as a theoretical idealization.

In practice, such a balanced superposition is difficult to engineer exactly in photonic or optomechanical platforms. However, it can be approximately realized by applying a weak coherent displacement operator to the vacuum state:
\begin{equation}
D(\alpha) = \exp\left( \alpha \hat{a}^\dagger - \alpha^* \hat{a} \right),
\end{equation}
which for small \( |\alpha| \ll 1 \) yields the expansion
\begin{equation}
D(\alpha)|0\rangle \approx |0\rangle + \alpha |1\rangle + \mathcal{O}(\alpha^2).
\end{equation}
For example, taking \( \alpha = 0.1 \) gives an unbalanced coherent superposition:
\begin{equation}
|0\rangle \rightarrow |0\rangle + 0.1 |1\rangle.
\end{equation}

Although this state is not equivalent to an ideal Hadamard rotation, it maintains quantum coherence between the vacuum and single-photon components and suffices to encode a relative phase shift in the interferometric sequence. While the interference \emph{visibility} is reduced compared to the balanced superposition, the resulting \emph{phase difference} \( \Delta \Phi \) remains well-defined and experimentally measurable. The Hadamard operation in our protocol should therefore be viewed as a conceptual tool for illustrating phase sensitivity, with weak coherent pulses providing a physically feasible route to implementing the required quantum control.

\subsection{Step 3: Conditional Adiabatic Evolution}

The mirror then evolves under two conditional Hamiltonians, depending on the TLS state. The radiation-pressure interaction couples the photon number \cite{Law:1995} $\hat{n}_{\text{ph}}$ to the mirror displacement $\hat{\xi}_1$ via
\begin{equation}
    \hat{H}_{\text{int}} = -\hbar g\, \hat{n}_{\text{ph}} \hat{\xi}_1.
\end{equation}
In branch A (TLS in state $|1\rangle$), the photon exerts a constant radiation pressure $F_{\text{rad},A} = \hbar g$, displacing the mirror. In branch B (TLS in $|0\rangle$), no such force acts.

Crucially, in the presence of gravitational waves, the two branches evolve against a background of spacetime curvature with distinct physical consequences. The \(|1\rangle\) branch, subject to radiation pressure, undergoes a time-dependent displacement, while the \(|0\rangle\) branch evolves passively. Both trajectories experience adiabatic modulation due to the gravitational-wave–induced time dependence of the trap parameters \(\omega_0(t)\) and \(Y_g(t)\). The complete time-dependent Hamiltonians governing the mirror motion in the two branches are:
\begin{align}
    \hat{H}_A(t) &= \frac{1}{2} (Z \hat{p}_1^2 + X(t) \hat{\xi}_1^2 + Y_g(t)(\hat{\xi}_1 \hat{p}_1 + \hat{p}_1 \hat{\xi}_1)) - \hbar g \hat{\xi}_1, \\
    \hat{H}_B(t) &= \frac{1}{2} (Z \hat{p}_1^2 + X(t) \hat{\xi}_1^2 + Y_g(t)(\hat{\xi}_1 \hat{p}_1 + \hat{p}_1 \hat{\xi}_1)).
\end{align}

The corresponding time-evolution operators are:
\begin{align}
    \hat{U}_A(T) &= \mathcal{T} \exp \left[ -\frac{i}{\hbar} \int_0^T dt' \, \hat{H}_A(t') \right], \\
    \hat{U}_B(T) &= \mathcal{T} \exp \left[ -\frac{i}{\hbar} \int_0^T dt' \, \hat{H}_B(t') \right],
\end{align}
where $\mathcal{T}$ denotes time-ordering.

The total state becomes:
\begin{align}
|\Psi(T)\rangle = \frac{1}{\sqrt{2}} \Big( 
&\hat{U}_B(T)|\psi(0)\rangle_m \otimes |0\rangle \nonumber \\
+\, &\hat{U}_A(T)|\psi(0)\rangle_m \otimes |1\rangle \Big).
\end{align}

%\begin{equation}
    %|\Psi(T)\rangle = \frac{1}{\sqrt{2}} \left( \hat{U}_B(T)|\psi(0)\rangle_m \otimes |0\rangle + \hat{U}_A(T)|\psi(0)\rangle_m \otimes |1\rangle \right).

    \subsection{Step 4: Second Weak Coherent Pulse and Interference}

Furthermore, in our tabletop model, a second Hadamard-like operation is applied to complete the Ramsey interferometry sequence:
\begin{align}
    |0\rangle &\rightarrow \frac{1}{\sqrt{2}} (|0\rangle + |1\rangle), \\
    |1\rangle &\rightarrow \frac{1}{\sqrt{2}} (|0\rangle - |1\rangle),
\end{align}
leading to the final entangled state:
\begin{equation}
    |\Psi(T)\rangle = \frac{1}{2} \left[ (\hat{U}_B + \hat{U}_A)|\psi(0)\rangle_m \otimes |0\rangle + (\hat{U}_B - \hat{U}_A)|\psi(0)\rangle_m \otimes |1\rangle \right].
\end{equation}

In practice, this second Hadamard-like transformation can again be approximated using a weak coherent pulse acting on the radiation field. Although the resulting superposition is unbalanced—e.g., \( |0\rangle \rightarrow |0\rangle + \alpha |1\rangle \)—the interference mechanism remains operational. The resulting phase shift \(\Delta \Phi = \Phi_A - \Phi_B\) manifests in the interference fringes, even though the contrast may be reduced relative to the ideal case.

Therefore, similar to Step 2, this second weak pulse serves as a physical realization of the Ramsey-type transformation needed to extract gravitationally induced phase differences. The use of a balanced Hadamard gate in the theory should be viewed as an analytically convenient simplification that captures the essential quantum interference phenomena enabled by weak coherent state control.

%\end{equation}

\subsection{Step 5: Readout and Phase Extraction}

Projecting onto the internal basis $\{|0\rangle, |1\rangle\}$, the probability of measuring the system in state $|0\rangle$ is given by
\begin{equation}
    P_0 = \frac{1}{2} \left[ 1 + \text{Re} \left( \langle \psi(0)|_m \hat{U}_B^\dagger \hat{U}_A |\psi(0)\rangle_m \right) \right].
\end{equation}

Under the assumption of adiabatic evolution—enabled by the low-frequency character of the gravitational wave and synchronized time modulation of the trap frequency—the time-evolution operators $\hat{U}_A$ and $\hat{U}_B$ remain unitary, and the motional states stay normalized:

\begin{align}
    \hat{U}_A (T)|\psi(0)\rangle_m &= e^{i \Phi_A} |\psi(T)\rangle_m, \\
    \hat{U}_B (T)|\psi(0)\rangle_m &= e^{i \Phi_B} |\psi(T)\rangle_m.
\end{align}
then:
\begin{equation}
    \langle \psi(0)| \hat{U}_B^\dagger \hat{U}_A |\psi(0)\rangle = e^{i(\Phi_A - \Phi_B)} = e^{i \Delta \Phi}.
\end{equation}

Hence, the final result for the interference signal becomes

\begin{equation}
    P_0 \approx \frac{1}{2} \left[ 1 + \cos(\Delta \Phi) \right],
\end{equation}
where $\Delta \Phi = \Phi_A - \Phi_B$ is the total accumulated phase difference.

\subsection{Total Phase and Quantum Signature}

The total phase difference accumulated between the two interferometric branches,
\begin{equation}
    \Delta \Phi_{tot} = \Phi_A - \Phi_B,
    \label{fg}
\end{equation}
generally comprises three types of contributions:
\begin{equation}
    \Delta \Phi_{tot} = \Delta \Phi_{\text{dyn}} + \Delta \Phi_{\text{Berry}} + \Delta \Phi_{\text{AB}},
\end{equation}
where $\Delta \Phi_{\text{dyn}}$ is the dynamical phase, $\Delta \Phi_{\text{Berry}}$ is the adiabatic Berry phase, and $\Delta \Phi_{\text{AB}}$ is an Aharonov--Bohm--like geometric phase.

In our configuration, both branches experience identical gravitational wave amplitude $Y_g(t)$ and identical modulation of the trap frequency $\omega_0(t)$, leading to matched adiabatic evolution. Provided the mirror starts in the same quantum state in each branch, this symmetry ensures that the Berry phases cancel out:
\begin{equation}
    \Delta \Phi_{\text{Berry}} = \Phi_A^{\text{Berry}} - \Phi_B^{\text{Berry}} \approx 0.
\end{equation}

Thus, the net interferometric phase consists of two surviving terms:
\begin{itemize}
    \item A \textit{dynamical phase shift} originating from the radiation pressure force acting only in the branch $A$, which shifts the energy spectrum as
    \begin{equation}
        \Delta \Phi_{\text{dyn}} = \frac{1}{\hbar} \int_0^T \left[ -\frac{F_{\text{rad}}^2 Z^2}{2 \omega^2(t)} \right] dt,
    \end{equation}

Using the specific system parameters introduced earlier in equations \eqref{p} and \eqref{b}, this expression simplifies to
    \begin{equation}
        \Delta \Phi_{\text{dyn}} = - \frac{\pi F_{\text{rad}}^2}{\hbar m_0^2 \omega_0^2 \omega_g} \cdot \frac{1}{(1 - \varepsilon)^{3/2}}.
    \end{equation}

    For \(\varepsilon \ll 1\), the dynamical phase can be approximated as
    \begin{equation}
      \Delta \Phi_{\text{dyn}} \approx - \frac{\pi F_{\text{rad}}^2}{\hbar m_0^2 \omega_0^2 \omega_g} \left( 1 + \frac{2}{3} \varepsilon + \cdots \right).  
    \end{equation}

    \item An \textit{Aharonov–Bohm–like (AB-like) geometric phase}, induced by radiation-pressure-driven coherent displacement evolving periodically in a background of weak gravitational waves:
    \begin{equation}
        \Delta \Phi_{\text{AB}}= -\oint_C \frac{1}{2\hbar} a^2 \nabla_{\vec{R}} \left( \frac{Y_g}{Z} \right) \cdot d\vec{R},
    \end{equation}
    where $a(t) = F_{\text{rad}} Z / \omega^2(t)$ is the radiation-pressure-induced displacement.
\end{itemize}

Likewise, in the small-\(\varepsilon\) limit, this simplifies to

\begin{equation}
    \Delta \Phi_{\mathrm{AB}}= - \frac{4 \pi \omega_g \chi_0 \epsilon_+ F_{\mathrm{rad}}^2 Z}{\hbar \Omega^2 \sqrt{\varepsilon}} + \mathcal{O}(\varepsilon).
\end{equation}

Although both phase contributions scale with \( 1/\hbar \), the radiation pressure itself arises from a quantized interaction, \( F_{\text{rad}} = \hbar g \), ensuring that the total phase remains linear in \( \hbar \). This confirms that both the dynamical and geometric phases are fundamentally quantum in origin. However, the dynamical phase depends explicitly on the detailed form of the system's time evolution and lacks any associated geometric structure—it reflects no underlying curvature or holonomy in parameter space. In contrast, the AB-like phase preserves its geometric character, manifesting as a contour integral over the trajectory in parameter space and encoding information about the global, path-dependent structure of the modulation.

\section{Interferometric Isolation of the Geometric Phase}

In our Ramsey-type setup, the total interferometric phase consists of two distinct contributions: a \textit{dynamical phase shift} \( \Delta \Phi_{\text{dyn}} \), arising from energy level shifts induced by radiation pressure, and a \textit{geometric (Aharonov–Bohm–like) phase shift} \( \Delta \Phi_{\text{AB}} \), generated by the interplay between gravitational wave (GW) strain and the quantum modulation path. Importantly, these two components exhibit different symmetry properties under time reversal of the modulation profile: while the dynamical phase remains invariant, the geometric phase changes sign due to its dependence on the direction of traversal in parameter space,
\begin{equation}
    \Delta \Phi_{\text{AB}} \longrightarrow -\Delta \Phi_{\text{AB}} \quad \text{under} \quad \vec{R}(t) \to \vec{R}(T - t).
\end{equation}
This contrast provides a practical route to extract the geometric phase. By comparing interferometric outputs from two otherwise identical sequences—one executed with forward modulation and the other with its time-reversed counterpart—the geometric contribution can be cleanly isolated via antisymmetric differencing.

To further disentangle the GW-induced geometric phase from the total signal, we consider two  configurations guided by Eq.~\ref{fg}:

\begin{itemize}
    \item \textbf{Baseline (no GW coupling):}
    \begin{equation}
        \Delta \Phi_{\text{tot}}(0) = \Delta\Phi_{\text{dyn}},
    \end{equation}
    
    \item \textbf{With GW coupling:}
    \begin{equation}
        \Delta \Phi_{\text{tot}}(\chi_0) = \Delta\Phi_{\text{dyn}} + \Delta\Phi_{\text{AB}}(\chi_0).
    \end{equation}
\end{itemize}
The resulting phase shift difference,
\begin{equation}
    \Delta \Phi \equiv \Delta\Phi_{\text{tot}}(\chi_0) - \Delta\Phi_{\text{tot}}(0) = \Delta\Phi_{\text{AB}}(\chi_0),
\end{equation}
provides a clean experimental handle on the gravitationally induced geometric phase, thereby extracting quantum features of spacetime curvature that are otherwise obscured by dynamical contributions.

Furthermore, to quantify the signal’s observability, we define the dimensionless phase ratio
\begin{align}
    R_{\text{GW}} 
    &:= \left| \frac{\Delta\Phi_{\text{AB}}}{\Delta\Phi_{\text{dyn}}} \right| \notag \\
    &= \left| \frac{\Delta\Phi_{\text{tot}}(\chi_0) - \Delta\Phi_{\text{tot}}(0)}{\Delta\Phi_{\text{tot}}(0)} \right| \notag \\
    &= \frac{4 \omega_g^2 \chi_0 \epsilon_+}{\varepsilon^{3/2}}.
\end{align}
%\begin{equation}
    %R_{\text{GW}} := \left| \frac{\Delta\Phi_{\text{AB}}}{\Delta\Phi_{\text{dyn}}} \right| 
   % = \left| \frac{\Delta\Phi_{\text{tot}}(\chi_0) - \Delta\Phi_{\text{tot}}(0)}{\Delta\Phi_{\text{tot}}(0)} \right|
   % = \frac{4 \omega_g^2 \chi_0 \epsilon_+}{\varepsilon^{3/2}}.
%\end{equation}
This ratio captures the relative strength of the geometric AB-ike phase independent of the absolute phase magnitude. Even if both \( \Delta\Phi_{\text{dyn}} \) and \( \Delta\Phi_{\text{AB}} \) may fall below the single-shot detection threshold, the dimensionless quantity \( R_{\text{GW}} \) can still enter a detectable regime through repeated measurements, long coherence times, or interference-enhanced protocols. This strategy parallels established methods in Berry-phase experiments, where small geometric phase shifts are revealed using differential and cyclic modulation techniques.

\begin{figure}[ht]
    \centering
    \includegraphics[width=0.48\textwidth]{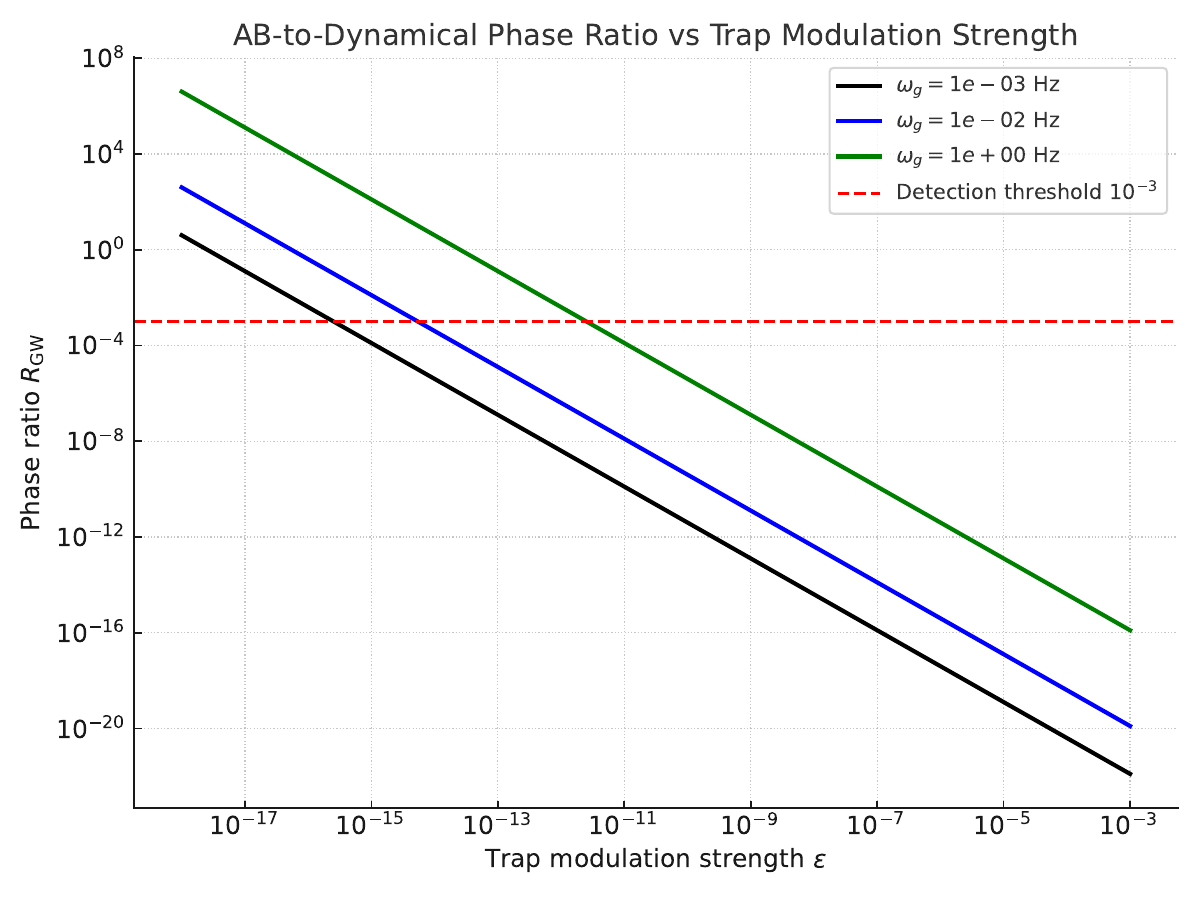}
    \caption{
       Log–log plot of the dimensionless phase ratio \( R_{\text{GW}} = |\Phi_{\text{AB}} / \Phi_{\text{dyn}}| \) versus trap modulation strength \( \varepsilon \), shown for three gravitational wave frequencies \( \omega_g = 10^{-3}, 10^{-2}, 1 \, \text{Hz} \). A horizontal red dashed line marks a conservative detection threshold \( R_{\text{GW}} = 10^{-4} \). For fixed GW strain \( \chi_0 \sim 10^{-21} \), the AB phase becomes relatively stronger in tighter traps (smaller \( \varepsilon \)) and at higher frequencies.
}
    
    \label{fig:phase-ratio}
\end{figure}

In Fig.~\ref{fig:phase-ratio}, we plot the dimensionless phase ratio \( R_{\text{GW}} \equiv |\Phi_{\text{AB}} / \Phi_{\text{dyn}}| \) as a function of the modulation strength \( \varepsilon \), for three representative gravitational wave frequencies. A detection threshold of \( R_{\text{GW}} \sim 10^{-4} \) is shown for reference \cite{Ramsey1950}. Notably, even for ultra-low-frequency gravitational waves (e.g., \( \omega_g \sim 10^{-3} \, \mathrm{Hz} \))—a range relevant for next-generation low-frequency detectors~\cite{Heuer2020}—a sufficiently small modulation strength (\( \varepsilon \sim 10^{-17} \)) can raise the AB-like geometric-to-dynamical phase ratio into the experimentally accessible regime.

\section{Conclusion}

We have developed a quantum-optomechanical framework for detecting gravitational waves (GWs) through purely quantum geometric phases, offering a conceptually distinct alternative to conventional displacement-based detection strategies. While prior studies—including earlier work by one of us~\cite{nandi2023berry}—demonstrated that time-dependent spacetime curvature can induce Berry phases in parametrically modulated quantum oscillators, such analyses remained confined to idealized, closed systems. They did not address the influence of additional physical effects, such as radiation pressure, nor did they propose realistic detection protocols.

In this work, we have significantly extended this paradigm along several key directions. First, we constructed a physically grounded model that incorporates both GW-induced parametric squeezing and radiation-pressure-induced coherent displacement. We showed that this setting gives rise to two distinct types of quantum phases: the well-known Berry phase associated with adiabatic evolution, and a previously unexplored Aharonov--Bohm--like (AB-like) geometric phase arising from cyclic displacement in the Hamiltonian's parameter space. While the Berry phase admits a classical analog in the form of the Hannay angle, the AB-like phase has no classical counterpart. Instead, it reflects a global, gauge-like memory of the system's full quantum trajectory—even though the classical path returns to its initial configuration.

Second, we derived closed-form expressions for both phase contributions and analyzed their scaling with experimentally tunable parameters, including gravitational wave amplitude, trap frequency modulation, and radiation pressure strength. Importantly, we proposed a Ramsey-type optomechanical interferometric scheme capable of isolating the AB-like phase by canceling both the Berry and dynamical components under symmetric evolution. This constitutes a realistic and experimentally accessible pathway for detecting purely quantum gravitational signatures in mesoscopic systems operating near their motional ground state.

Taken together, these results establish a novel and viable approach for sensing gravitational waves using phase-coherent quantum probes. Our findings illustrate that quantum evolution in curved spacetime can retain nontrivial geometric information—even in regimes where classical observables such as displacement vanish or are dominated by noise. The proposed protocol capitalizes on the quantum sensitivity of coherent mesoscopic wavepackets to background curvature and demonstrates how interferometric techniques, long used in classical GW detection, can be reimagined in the quantum domain to probe new physics.

%It is worth emphasizing that, unlike classical detection methods relying on measurable displacements, our proposal leverages the coherence properties of mesoscopic quantum systems. This makes it especially well-suited to probing low-frequency gravitational wave backgrounds, where adiabatic evolution enables geometric phase accumulation without exciting the system out of its ground state. The central role played by the center-of-mass quantum evolution is physically well-motivated: the translation-invariant structure of the effective Hamiltonian ensures that the COM mode remains decoupled from internal vibrational or thermal modes, preserving phase coherence over the duration of the interaction.

While the recent experimental observation of a gravitational Aharonov--Bohm effect~\cite{overstreet2022gravityAB} demonstrates the sensitivity of quantum phases to static Newtonian curvature via proper-time differences, our framework offers a fundamentally distinct viewpoint. We focus instead on dynamical gravitational fields—specifically, low-frequency gravitational waves (LFGWs)—and show that even in the absence of proper-time differentials or measurable displacements, purely quantum geometric phases can emerge from the cyclic evolution of the Hamiltonian in parameter space. In contrast to the Overstreet experiment, which isolates classical deflection from phase accumulation, our model uncovers nonclassical memory effects encoded in the full quantum trajectory. These effects are experimentally accessible via Ramsey-type optomechanical interferometry, and they reflect genuinely dynamical features of the background spacetime associated with LFGWs.

The practical detectability of the AB-like phase is further illustrated in Fig.~\ref{fig:phase-ratio}, where we plot the dimensionless phase ratio \( R_{\text{GW}} \) as a function of the trap modulation strength \( \varepsilon \) for representative gravitational wave frequencies. This plot highlights the regimes where the geometric phase dominates the signal and enters the observable threshold, even for ultra-low frequency GWs \cite{Graham2016ulfgw}. Notably, the AB-like contribution becomes increasingly prominent for tighter traps (smaller \( \varepsilon \)) and higher modulation frequencies, offering concrete experimental guidance for optimizing detection sensitivity. This reinforces our claim that mesoscopic optomechanical platforms are well-positioned to access new gravitational signatures through quantum phase readout, even when classical displacement signals are suppressed.

Looking ahead, an important open question is whether the AB-like phase identified here persists—or takes on new features—in the presence of quantized gravitational wave (graviton) backgrounds~\cite{PhysRevD.109.044009}. Such scenarios could provide valuable insights into the microscopic structure of quantum spacetime \cite{Herceg:2023zlk} and offer potential observational signatures of graviton-mediated interactions. Detecting any effect arising from quantized gravitational degrees of freedom remains a formidable challenge, both conceptually and experimentally. Nonetheless, we anticipate that continued progress in ground-state-cooled optomechanical systems, along with advances in coherent quantum control, may eventually make it possible to access these intrinsically quantum geometric phase effects in the laboratory—opening a promising new window into the quantum nature of gravity~\cite{Belenchia2019tabletop}.

\begin{acknowledgments}
PN acknowledges support from the National Institute for Theoretical and Computational Sciences (NITheCS) through the Rector’s Postdoctoral Fellowship Program (RPFP). He is grateful to Prof. Yin-Zhe Ma for the opportunity to present a series of lectures on gravity-induced geometric phases to the astrophysics group at Stellenbosch University. He also thanks Prof. Soumen Mondal for helpful discussions and clarifications related to this work. We are especially grateful to Prof. M. V. Berry for his insightful comments and helpful suggestions on improving the manuscript.

\end{acknowledgments}

\bibliographystyle{apsrev4-2}
\bibliography{gw_phases}

\end{document}